\documentclass[aps,prb,floats,showpacs,twocolumn]{revtex4}
\usepackage{graphicx,soul,color,mathrsfs,wrapfig,amsmath}

\renewcommand{\bbox}[1]{\raisebox{\depth}{#1}}

\newcommand{\cbox}[1]{\raisebox{-0.5\height}{\bbox{#1}}}

\sethlcolor{yellow}

\newcommand{\bra}[1]{\langle #1 |}
\newcommand{\ket}[1]{| #1 \rangle}
\newcommand{\mvec}[1]{\mathbf{#1}}

\newcommand{\be}{\begin{equation}}
\newcommand{\ee}{\end{equation}}

\def \ua{{\uparrow}}
\def \da{{\downarrow}}
\def \be{\begin{equation}}
\def \ee{\end{equation}}
\def \ba{\begin{array}}
\def \ea{\end{array}}
\def \bea{\begin{eqnarray}}
\def \eea{\end{eqnarray}}
\def \nn{\nonumber}

\def \half{{1\over 2}}

\def \bJ{{\bf J}}
\def \bA{{\bf A}}

\def \bv{{\bf v}}

\def \bk{{\bf k}}

\def \bx{{\bf x}}

\def \e{{\epsilon}}

\def \a{{\alpha}}

\def \g{{\gamma}}
\def \D{{\Delta}}
\def \d{{\delta }}

\def \s{{\sigma}}

\def \e{{\epsilon}}

\def \nd{{^{\vphantom{\dagger}}}}
\def \yd{^\dagger}
\def \av#1{{\langle#1\rangle}}
\def \ket#1{{\,|\,#1\,\rangle\,}}
\def \bra#1{{\,\langle\,#1\,|\,}}

\begin{document}
\title{Enhancement of the superconducting transition temperature in cuprate heterostructures}
\author{Lilach Goren}
\affiliation{Department of Condensed Matter Physics, The Weizmann Institute of Science, 76100 Rehovot (Israel)}
\author{Ehud Altman}
\affiliation{Department of Condensed Matter Physics, The Weizmann Institute of Science, 76100 Rehovot (Israel)}
\date{\today}
\begin{abstract}
Is it possible to increase $T_c$ by constructing cuprate heterostructures, which combine the high pairing energy of underdoped layers with the large carrier density of proximate overdoped layers? We investigate this question within a model bilayer system using an effective theory of the doped Mott insulator. Interestingly, the question hinges on the fundamental nature of the superconducting state in the underdoped regime. Within a plain slave boson mean field theory, there is absolutely no enhancement of $T_c$. However, we do get a substantial enhancement for moderate inter-layer tunneling when we use an effective low energy theory of the bilayer in which the effective quasiparticle charge in the underdoped regime is taken as an independent phenomenological parameter. We study the $T_c$ enhancement as a function of the doping level and the inter-layer tunneling, and discuss possible connections to recent experiments by Yuli {\em et al.} [Phys. Rev. Lett. {\bf 101}, 057005 (2008)]. Finally, we predict a unique paramagnetic reduction of the {\em zero temperature phase stiffness} of coupled layers, which depends on the difference in the current carried by quasiparticles on the two types of layers as $(\bJ_1-\bJ_2)^2$.
\end{abstract}
 \maketitle
\section{Introduction}\label{section:intoduction}
There are strong indications\cite{Uemura,EmeryKivelson,Corson,Ong} that the superconducting transition temperature of underdoped cuprate materials is limited only by their small superfluid density, while the pairing scale is very high. This understanding has raised the hope that $T_c$ can yet be made substantially higher by clever design of materials.
In particular, a number of recent theoretical\cite{Kivelson,Erez,Maier} and experimental\cite{Yuli,BozovicPhysica,BozovicNature} studies explored the possible benefit in heterostructured materials combining metallic layers with layers
of underdoped cuprate material. The basic idea is simple; underdoped layers contribute a strong microscopic pairing interaction, whereas metallic layers provide a high density of charge carriers. But it could also go the other way. Namely, the metal destroys pairing in the underdoped layer without contributing much of its charge carriers. The question how much, if at all, such systems can actually enhance $T_c$ may require deeper knowledge of the nature of the superconducting state in the cuprates. Of particular importance in this respect is better understanding of the mechanisms that reduce the superfluid density with temperature and with proximity to the Mott insulating state.

In this paper we investigate the problem of $T_c$ enhancement within a bilayer model using an effective description of the doped Mott insulator. The model captures an interesting competition of effects, which we expect is rather general  to cuprate heterostructures and possibly inhomogeneous realizations of these materials.
In particular we discuss possible implications of our results to recent experiments in  ${\rm La_{2-x}Sr_xCu O_4}$  (LSCO)  bilayers\cite{Yuli}. We argue that experiments with heterostructures may shed new light on fundamental questions concerning the nature of superconductivity in cuprates.

\begin{figure}[t]
\begin{center}
\includegraphics{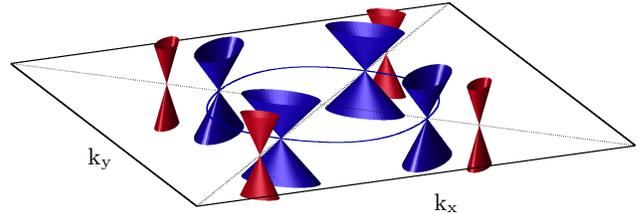}
\end{center}
\caption{Illustration of the two dimensional low energy dispersion of a bilayer consisting of a d-wave superconducting layer and a nominally metallic layer. The four red surfaces mark the original Dirac cones of the superconducting layer. The blue surfaces are the Dirac cones induced by the proximity effect on the metallic layer. The blue curve denotes the original Fermi surface of that layer.}
\label{fig:NodalCones}
\end{figure}

The essential physics that determines $T_c$ of the bilayer is most clearly illustrated within the effective low energy theory of $d$-wave superconductors\cite{WenLee,Millis,IoffeMillis}. The superfluid stiffness follows a linear temperature dependence at low temperatures $\rho_s(T)\approx \rho_s(0) -B T$ due to thermal excitation of quasiparticles at the Dirac nodes. The slope $B$ is inversely proportional to the magnitude of the $d$-wave gap. This expression implies a crude estimate of $T_c$ in a two dimensional system $T_c\sim \rho_s(0)/(B+2/\pi)$, which is the point where the criterion for a Kosterlitz-Thouless transition is satisfied ($\rho_s(T_c)=2T_c/\pi$).
Crucially, the transition temperature depends on the zero temperature stiffness, but also on the quasiparticle gap via the slope $B$.

Coupling a layer of $d$-wave superconductor to a normal layer produces a $d$-wave proximity gap, which protects a super-flow of electrons in the normal layer. Fig.~\ref{fig:NodalCones} is an illustration of the resulting low energy spectrum of the bilayer system. While the carrier density of the underdoped layer is  small, proportional to hole doping of the Mott insulator, that of the normal layer is much larger and is of the order of the total electron density. Consequently, the zero temperature superfluid stiffness is huge, consisting of the contributions from the two layers and is clearly dominated by the carrier density of the normal layer $\rho_s(T=0)\approx n_{s1}/m_1^\star+n_{s2}/m_2^\star$.
On the other hand the reduction of the stiffness with temperature is now much steeper than for a single layer. This is because  it is dominated by thermal excitation of quasiparticles at the nodes of the d-wave proximity gap, which is smaller than the pairing gap in the underdoped layer. At weak coupling between the layers, the proximity gap is given by $\D^{\rm prox}_\bk\approx \D_\bk ({\tilde t}_\perp/E_1)^2$,
where ${\tilde t}_\perp$ is an effective inter-layer tunneling and $E_1$ is a larger energy scale determined by the mismatch of the two Fermi surfaces. The resulting linear slope
of the stiffness with temperature, is given to leading order in the small parameter ${\tilde t}_\perp/E_1$ by $B_2\approx(2\ln 2/ \pi)  \left(v_{F 2}/ v_\D \right)(E_1/{\tilde t}_\perp)^2$. Here $v_{F 2}$ is the Fermi velocity of the normal layer and $v_\D$ the slope of the pairing gap of the underdoped layer at the node. This should be compared with the smaller slope in a pure underdoped material $B_1= \a^2(2\ln 2/ \pi)  \left(v_{F 1}/ v_\D \right)$, where $\a$ is the effective electric charge carried by a current of quasiparticles. Thus the question of $T_c$ enhancement in the bilayer hinges on the competition between the increased zero temperature stiffness and reduced quasiparticle gap compared to the pure underdoped material.

To understand the full dependence of $T_c$ enhancement on doping level and bilayer coupling one must go beyond these leading order estimates. To do this we draw on the basic framework of slave boson mean field theory (SBMFT)\cite{SBMFT1,SBMFT2,SBMFT3}. We also derive an effective semi-phenomenological theory of the bilayer, along the lines of Ref. [\onlinecite{IoffeMillis}], which keeps the spirit of SBMFT while avoiding some of its peculiarities. The resulting phase diagram shows significant enhancement of $T_c$ for moderate inter-layer tunneling over a wide range of doping levels. As a function of inter-layer tunneling, $T_c$ increases at first, but reaches a maximum at an optimal value of $t_\perp$. The main factor in setting the
optimal  coupling is the anti-proximity effect on the pairing gap of the underdoped layer.

We find another, somewhat more subtle, anti-proximity effect, which affects the zero temperature superfluid stiffness. In addition to the usual diamagnetic response, there is a {\em paramagnetic} correction at zero temperature due to mixing of quasiparticle wave-functions between the two layers. Accordingly, the zero temperature superfluid stiffness is smaller than the independent contributions of the two layers, $n_{s1}/m_1^\star+n_{s2}/m_2^\star$, by a term proportional to $(J_1-J_2)^2$, where $J_l=\a_l v_{F l}$ is the current carried by a quasiparticle located on layer $l$. We argue that measurements of this effect can lend insights into a long standing problem concerning the so-called quasiparticle charge\cite{Millis,LeeReview}, or current carried by a quasiparticle in the cuprates.

The rest of this paper is organized as follows. In section \ref{section:model} we define the microscopic model which serves as the basis for our theoretical analysis. In section \ref{section:SBMFT} we use self consistent slave boson mean field theory to obtain the temperature dependent phase stiffness and a phase diagram of the bilayer model. In section \ref{section:effectiveH} we derive a low energy effective theory of the response to an external vector potential starting the from the slave boson formulation. We then generalize the low energy theory to include renormalized parameters for the zero temperature superfluid stiffness and the effective quasiparticle charge. In section \ref{subsec:Tdepstiffness} we use the semi-phenomenological theory to predict $T_c$ enhancement in a
 putative LSCO bilayer composite system. In section \ref{subsec:zeroT} we use the effective theory to derive the paramagnetic correction to the zero temperature stiffness. Finally in section \ref{sec:summary} we summarize our main conclusions and discuss possible implications to recent experiments.

\section{The Model}\label{section:model}
Our starting point for theoretical investigation is the following model of a bilayer system:
\begin{eqnarray}\label{H}
H & = & H_1+H_2+H_\perp \nonumber\\
H_1 & = &-t_1\sum_{\langle ij\rangle
\sigma}P[c_{i\sigma}^\dagger c_{j\sigma} +h.c.]P-
(\e_0+\mu)\sum_{i\s}c\yd_{i\s}c\nd_{i \s}  \nonumber\\
&&+  J\sum_{\langle
ij\rangle}[\mvec{S}_{i}\cdot \mvec{S}_{j}-\frac{1}{4}n_i n_j]+\ldots\nonumber\\
H_2 & = &-t_2\sum_{\langle ij\rangle-
\sigma}[d_{i\sigma}^\dagger d_{j\sigma} +h.c.] -\mu\sum_i d\yd_{i\s }d\nd_{i\s }\nonumber\\
H_\perp & = & -t_\perp \sum_{i\sigma} [c_{i\sigma}^\dagger d_{i\sigma} +h.c].
\end{eqnarray}
Here, the normal (highly overdoped) layer is modeled by the Hamiltonian $H_2$ of non interacting Fermions $d\yd_{i\s}$ on the square lattice. The underdoped layer on the other hand is modeled by the effective $t-J$ Hamiltonian $H_1$, which takes into account the proximity of the Mott insulating state.
$P$ is the projection on the low energy subspace with no doubly occupied sites, and the dots represent possible additional terms. The energy offset $\e_0$ is the single layer chemical potential that would set the correct hole doping of the underdoped layer in absence of inter-layer coupling $H_\perp$. When the two layers are coupled by $H_\perp$ they of course must share a common chemical potential $\mu$, which in general leads to charge redistribution between the layers.

\section{Mean Field phase diagram}\label{section:SBMFT}
In this section we obtain the phase diagram of the bilayer model using
the slave boson mean field theory (SBMFT)\cite{SBMFT1,SBMFT2,SBMFT3}. This is the simplest theory, that gives a BCS like superconductor with a large Fermi surface but with low superfluid density, which scales as the hole doping. Thus for a single layer $T_c$ is controlled by the zero temperature superfluid stiffness, rather than by the pairing gap.

\subsection{Slave boson mean field theory for a bilayer}
Before presenting the bilayer calculation, let us briefly review the standard slave boson approach for a single underdoped layer. The electron creation operator is represented as a composite of a fermionic spinon and a bosonic holon operator $c^\dag_{i\sigma}=b_i f^\dag_{i\sigma}$. The redundancy of representation is
removed by the local constraint $b\yd_i b\nd_i+\sum_\s f\yd_{i\s} f\nd_{i\s}=1$, which can be implemented exactly by a $U(1)$ gauge field. The core approximation of the mean field solution is that at least in the superconducting phase, both the holon and the gauge field are condensed. This allows to replace the operator $b_i$ by the number $[2x/(1+x)]^{1/2}\equiv\sqrt{\delta }$ and implement the constraint only on the average\cite{SBMFT1}. At this stage
  the approximate Hamiltonian $H'(x)$ is written in terms of the fermion spinon operators only, and acts in an unrestricted Hilbert space, however it is still quartic. The second approximation consists of a standard mean field solution of $H'$, whereby one seeks the best quadratic approximation to it of the form
\bea\label{HSB:extField}
H_0 &=&-t_1 \delta \sum_{\av{ij}}\left(e^{ieA_{ij}} f\yd_i f\nd_j+h.c.\right)-\e_0\sum_i f\yd_i f\nd_i\nonumber\\
&&+\sum_{\bk\s}( \D_\bk f\yd_{\ua,\bk} f\yd_{\da,-\bk} h.c.) -\sum_{\bk\s}
\chi_\bk f\yd_{\s \bk} f\nd_{\s \bk},
\eea
where $\D_\bk =\D(\cos k_x- \cos k_y)$ and $\chi_\bk=\chi(\cos k_x+ \cos k_y)$. Here we introduced a coupling to an external vector potential through the phases $eA_{ij}$, which will later facilitate calculation of the superfluid density. Note that the electromagnetic vector potential  couples only to the charged holon field. Condensation of the holon leads to effective coupling to the fermion field in the kinetic energy term.
The parameters $\D$ and $\chi$ are determined using a general thermodynamic variational principle by minimization of
\be
F_{0}+\av{H'-H_0}_0,
\ee
$F_{0}$ is the free energy implied by the trial Hamiltonian $H_0$, and $\av{}_0$ denotes a thermal average generated by $H_0$. The chemical potential $\e_0$ is determined by resolution of the average constraint equation:
$\sum_\s\av{f\yd_{\s i}f\nd_{\s i}}+x=1$.

We now move on to include the inter-layer coupling.
At this point charge can be redistributed between the layers, changing the doping levels of the two layers from $x$ and $y$ in absence of the coupling to ${\tilde x}$ and ${\tilde y}$.  The quadratic inter-layer tunneling Hamiltonian is given by
\be
H_{\perp 0}=-t_\perp\sqrt{{\tilde \delta }} \sum_{i\sigma} [f_{i\sigma}^\dagger d_{i\sigma} +h.c],
\ee
where ${\tilde \d}=\sqrt{2{\tilde x}/(1+{\tilde x})}$. The quadratic bilayer (variational) Hamiltonian in momentum space is then
\begin{equation}\label{Hmf}
H_{MF}= \sum_{\mvec{k}}[\xi_{1\mvec{k}}+\xi_{2\mvec{k}}  + \Psi_\mvec{k}^\dagger h_\mvec{k} \Psi_\mvec{k}],
\end{equation}
where $\Psi_{\mvec{k}}^\dagger = (f_{\mvec{k}\uparrow}^\dagger , f_{-\mvec{k}\downarrow},d_{\mvec{k}\uparrow}^\dagger , d_{-\mvec{k}\downarrow})$ and
\begin{equation}\label{Hmatrix}
h_{\mvec{k}} =
\begin{pmatrix}
\xi_{1\mvec{k}}  &\Delta_\mvec{k}  & \tilde{t}_\perp& 0 \\
\Delta_\mvec{k} & -\xi_{1,-\mvec{k}} &0 & -\tilde{t}_\perp \\
\tilde{t}_\perp & 0 &  \xi_{2\mvec{k}}  &0 \\
0 & -\tilde{t}_\perp  & 0 &  -\xi_{2,-\mvec{k}}
 \end{pmatrix}
\end{equation}
with $\tilde{t}_\perp\equiv t_\perp \sqrt{{\tilde \delta }} $.
In the absence of external fields
\begin{eqnarray}
\xi_{1\mvec{k}}&=& -(2{\tilde \delta } t_1+\chi)(\cos{k_x} + \cos{k_y})-\mu \nonumber\\
\xi_{2\mvec{k}}&=& -2 t_2(\cos{k_x} + \cos{k_y})-\mu+\epsilon_0\nonumber\\
\Delta_{\mvec{k}}&=&\Delta(\cos{k_x} - \cos{k_y}).
\end{eqnarray}
The parameters $\D$ and $\chi$ can be determined again by solving the variational equations, supplemented by the two number equations for the additional unknowns
$\tilde x$ and $\mu$:
\bea
\av{n_f}&=&1-{\tilde x}\nonumber\\
\av{n_d}&=&1-{\tilde y}=1-y+({\tilde x}-x).
\eea
Clearly, a proximity gap will be induced in the normal layer due to the coupling
with the underdoped superconducting layer. The charge carriers in the second layer will then contribute to the superfluid density.

\begin{figure}
\begin{center}
\includegraphics{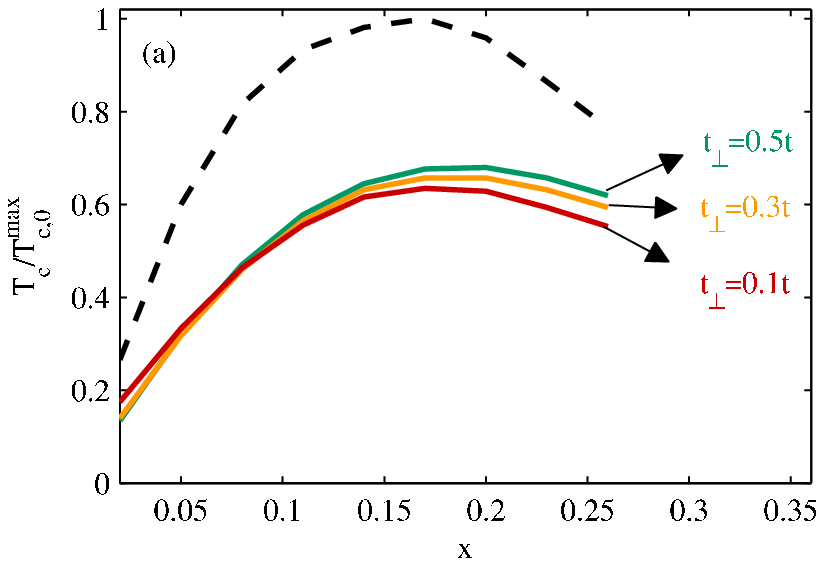}
\includegraphics{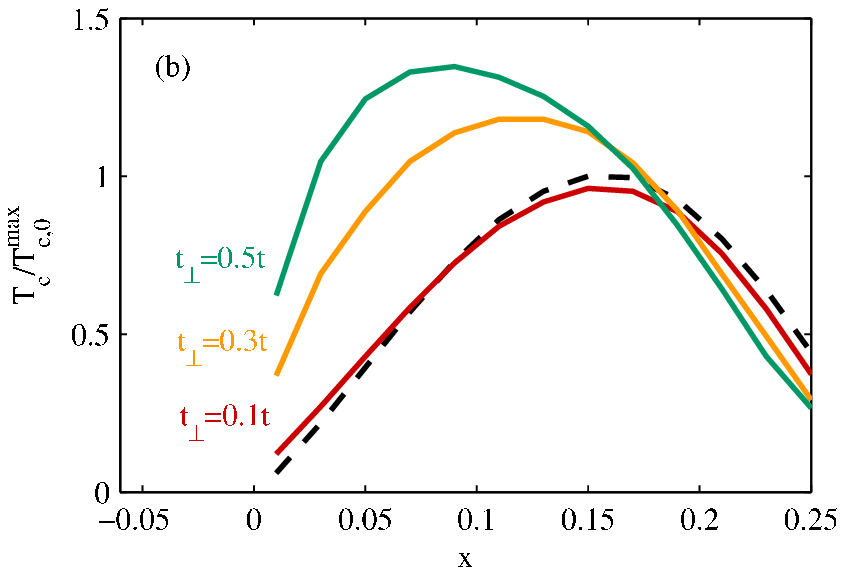}
\caption{{\em Phase diagram of the bilayer system from microscopic theory}. The critical temperature $T_c$ (normalized by its maximal value for two identical underdoped layers, $T_{c,0}^{\rm max}$) vs. doping of the underdoped layer. The doping level of the metallic layer is $y=0.35$. The dashed line is the result for two identical underdoped layers. (a) "Bare" SBMFT calculation. No enhancement of $T_c$. (b) SBMFT with renormalized quasiparticle charge of $\alpha_1=0.5$ in the underdoped layer. Maximal $T_c$ enhanced by $\sim 40\%$ in the heterostructure. Optimal doping shifted down, consistent with experiment of Yuli et. al. [\onlinecite{Yuli}].}
\label{fig:Tcvsx}
\end{center}
\end{figure}

\subsection{Superfluid density and $T_c$}\label{section:sfstiffness}
We shall obtain the critical temperature of the bilayer by computing the temperature dependent superfluid stiffness
\begin{equation}\label{rhos}
\rho_s(T)=\frac{1}{\Omega}\frac{\partial^2F}{\partial A^2}\bigg|_{A=0}.
\end{equation}
Here $F$ is the free energy, $A$ is an externally applied transverse vector potential and $\Omega$ is the volume of the system. The critical temperature is then given by the condition for a Kosterlitz-Thouless transition $\rho_s(T_c)=(2/ \pi)T_c$.

The dominant contribution to the reduction of superfluid stiffness at low temperatures in a d-wave superconductor is the paramagnetic response due to thermally excited quasiparticles in the gap nodes\cite{WenLee}.
This leads to a linear decrease of $\rho_s$ with $T$.
Because the paramagnetic response amounts to a current-current correlator, the slope $d\rho_s/dT$ is proportional to the square of the effective electric charge carried by a current of quasiparticles. Here we encounter a possible pitfall of the mean field theory. Within plain SBMFT, the quasiparticle charge is proportional to the doping $x$. However, experiments seem to point to a fairly doping independent value of this parameter\cite{panagopoulos:1999,Lemberger}. The experimental results can be reproduced by an effective theory, which maintains the spirit of SBMFT, but assigns a phenomenological value to the quasiparticle charge\cite{IoffeMillis}.

In our analysis of the two layer system we consider the bare SBMFT as well as a theory with a phenomenological quasiparticle charge renormalization.
The main result of the mean field calculation is a phase diagram of the bilayer heterostructure.

Fig. \ref{fig:Tcvsx}(a)
displays $T_c$ as a function of the doping $x$ computed using the bare slave boson theory for various values of the inter-layer tunneling $t_\perp$. No enhancement of $T_c$ relative to a pair of identical layers is found for the relevant range of parameters  $0.2<J<0.7$ and $0<t_\perp<1$.

In marked contrast, we do find
a significantly enhanced $T_c$ in a modified SBMFT which allows for a phenomenological quasiparticle charge renormalization independent of the doping. To compute the second order response (\ref{rhos}) to an external vector potential $\bA$, the mean field  Hamiltonian (\ref{HSB:extField}) is expanded to second order in $A_{ij}$
\begin{equation}
H(\bA)=H(0)-\sum_{\langle ij \rangle}  j_{ij}A_{ij} + \half \sum_{\langle ij \rangle} k_{ij} A_{ij}^2
\end{equation}
with the paramagnetic current operator $ j_{ij}= it_1\delta e \sum_\sigma (f_{i\sigma}^{\dag}f_{j\sigma}-f_{j\sigma}^{\dag}f_{i\sigma})$ and the average kinetic energy per bond $k_{ij}=- t_1\delta  e^2\sum_\sigma (f_{i\sigma}^{\dag}f_{j\sigma}+f_{j\sigma}^{\dag}f_{i\sigma})$. In the plain SBMFT approach the charge in the current operator $j_{ij}$ is renormalized by a factor $\delta \propto x$. It is this renormalization that leads to a strong doping dependence of the slope $d\rho_s/d_T$, at low temperatures\cite{WenLee}, which disagrees with experiments\cite{panagopoulos:1999,Lemberger}. As a possible cure of this artifact within the microscopic theory, we replace the factor $\delta $ in $j_{ij}$ by a doping independent number $\alpha$, which in principle should be determined experimentally. This is equivalent to introducing an effective quasiparticle charge of magnitude $\alpha e$ to all physical properties involving the quasiparticle current, as suggested in Refs. \onlinecite{WenLee} and \onlinecite{IoffeMillis}. In our model we apply this renormalization only to the underdoped layer since the normal layer is approximated simply by non interacting fermions. We choose a value of $\alpha\approx 0.5$, which reproduces the dome-shaped $T_c(x)$ phase diagram with $T_c/t_1$ having the right order of magnitude. For this renormalized value we obtain enhancement of $T_c$ of the bilayer compared to two identical underdoped layers with the same $\alpha$.

 The result of this calculation is presented in Fig. \ref{fig:Tcvsx}(b).
The optimal doping level of the underdoped layer of the heterostructure is seen to be around $x=0.1$, which is well below the optimal doping of the single layer and consistent with the result of recent experiments \cite{Yuli}.

The enhancement of $T_c$ stems from the combination of a large carrier number donated by the normal layer and a large pairing gap induced by the proximate underdoped layer. Because the proximity gap is smaller than the original pairing gap, the reduction of stiffness with temperature is also steeper in the heterostructure.
However, for the modified SBMFT with renormalized quasiparticle charge, the increase in slope $d\rho_s/dT$ is not large enough to offset the enhanced zero temperature superfluid stiffness. By comparison, in the plain SBMFT the slope of stiffness versus temperature for the pure underdoped material is much smaller because of the small quasiparticle charge. The increase in slope $d\rho_s/d_T$ on going from a pure underdoped material to a heterostructure is concomitantly more extreme. For this reason we see an enhancement of $T_c$ only for SBMFT with renormalized quasiparticle charge.

We note that the computation of superfluid stiffness (\ref{rhos}) with both plain and modified SBMFT is carried out fully self consistently. Thus it captures non linear contributions to the temperature dependence of the superfluid stiffness (but of-course not contributions from phase fluctuations).

\begin{figure}
\begin{center}
\includegraphics{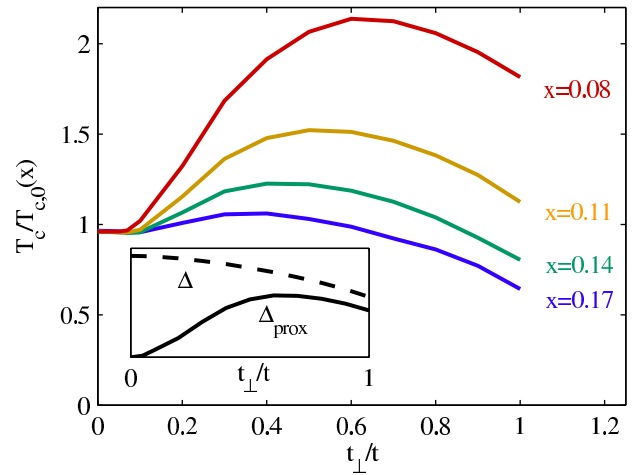}
\caption{{\em Optimal inter-layer tunneling.} Plotted is the critical temperature $T_c$ vs. the inter-layer coupling $t_\perp/t$ ($t=t_1=t_2$). Different curves correspond to different doping levels of the underdoped layer and all curves are normalized by the critical temperature of two underdoped layers of the same doping level, $T_{c,0}(x)$. The doping of the metallic layer is $y=0.35$. Inset: The self-consistent gap $\Delta$ and the proximity gap $\Delta_{\rm prox}$ as calculated from the bilayer energy spectrum, plotted vs. the inter-layer tunneling $t_\perp/t$ for doping $x=0.11$. This suggests that the optimal value of $t_\perp$ is determined by the point where the anti-proximity effect on the gap of the underdoped layer overtakes the proximity gap in the normal layer.}
\label{fig:Tcvstp}
\end{center}
\end{figure}

The degree of $T_c$ enhancement as a function of the inter-layer tunneling is plotted in Fig. \ref{fig:Tcvstp}. We note that the minimal inter-layer tunneling required to obtain such an enhancement ($t_\perp\sim t/4$) appears rather too large to serve as straightforward model of the bilayer experiment of Yuli et. al [\onlinecite{Yuli}]. This issue will be discussed further in section \ref{subsec:Tdepstiffness}.
Moving to still larger $t_\perp$ we observe a maximal enhancement of $T_c$ at $t_{\perp}\sim 0.5 t$ for which the enhancement may be as large as $40\%$.
The optimum value of $t_\perp$ occurs where the proximity gap becomes of order of the superconducting gap (see inset of Fig. \ref{fig:Tcvstp}). At this point the superconducting gap cannot increase any further and the anti-proximity effect of the normal layer on the superconducting one takes over.

 Before closing this section we point out another interesting effect in the response of the bilayer heterostructure as compared to the single layer (or a pair of identical layers). In the single layer mean field theory the paramagnetic response appears only due to quasiparticles at finite temperatures, whereas the zero temperature superfluid stiffness is independent of the effective quasiparticle charge. It is not so in the coupled bilayer heterostructure. In this case a quasiparticle carries a different current depending on whether it resides on the top or bottom layer. This allows redistribution of quasiparticles in the ground state in the presence of current and leads to a zero temperature paramagnetic contribution to the superfluid stiffness.

\begin{figure}
\begin{center}
\includegraphics{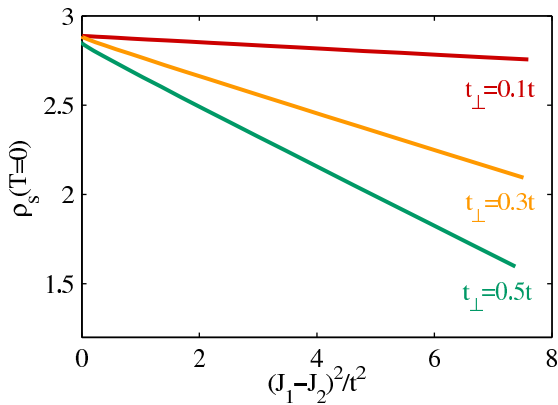}
\caption{{\em Paramagnetic reduction of the zero temperature stiffness.} $\rho_s(T=0)$ of the bilayer system (doping levels $x=0.11$ and $y=0.35$) is plotted against $(J_1-J_2)^2/t^2$, where $t=t_1=t_2$ is the bare in-layer tunneling and $J_l=\alpha_l v_{Fl}$ are the quasiparticle currents of the two layers. Here $v_{Fl}$ are the Fermi velocities at the nodes in the respective layers. We used the quasiparticle charge $\a_2=1$ for the normal layer and varied $\a_1$ of the underdoped layer.}
\label{fig:rho0vsJ}
\end{center}
\end{figure}

We will show in section \ref{subsec:zeroT}, that the reduction of the stiffness at zero temperature is proportional to $(\alpha_1v_{F1}-\alpha_2v_{F2})^2,$ where $\alpha_l$ and $v_{Fl}$ are the effective quasiparticle charge and the Fermi velocity in layer $l$. Note that $\alpha_l {\bv}_{Fl}={\bJ}_l$ is the current carried by a quasiparticle in layer $l$. Fig. \ref{fig:rho0vsJ} shows the zero temperature stiffness of the bilayer within the self consistent mean field calculation as a function of the quasiparticle charge renormalization. We see indeed that the correction is negative, and quadratic in $J_1-J_2$.


\section{Effective low energy theory}\label{section:effectiveH}
To clarify the mechanisms of $T_c$ enhancement and facilitate generalizations that are less dependent on a particular microscopic model it is worthwhile to derive a low energy effective theory for the bilayer system. This will be done perturbatively in the inter-layer coupling in subsection \ref{subsec:QPspectrum}. We shall also derive the effective coupling of the external field to the low energy Hamiltonian for the sake of computing the superfluid stiffness.
The low energy theory, with parameters extracted from bulk samples, will then be used to construct a phase diagram of the bilayer system for given values of the inter-layer coupling.

\subsection{Quasi-particle spectrum}\label{subsec:QPspectrum}
In the absence of inter-layer coupling, the lower diagonal block in the microscopic Hamiltonian (\ref{Hmatrix}) describes gapless particle and hole excitations near the Fermi surface of the metallic layer ($\xi^{(2)}_{\mvec{k}}\simeq0$). The primary effect of the coupling is to open a proximity gap in the metallic layer. This is captured nicely by the low energy effective Hamiltonian derived by second order degenerate perturbation theory. To this end it is convenient to rewrite the Hamiltonian (\ref{Hmatrix}) in terms of its $2X2$ blocks
\begin{equation}
h_{\mvec{k}} =
\begin{pmatrix}
h_1(\mvec{k}) & V\\
V^\dag &h_2(\mvec{k})
 \end{pmatrix}
 \end{equation}
 and treat $V$ as a perturbation. We note that since the effective inter-layer tunneling $\tilde{t}_\perp$ is proportional to $\sqrt{x}$, at sufficiently low doping levels a perturbative treatment may be justified even if the bare tunneling $t_\perp$ is not very small.

The low energy physics is dominated by excitations near the Fermi surface of the metallic layer (layer 2) and near the nodal points of the superconducting layer (layer 1).
 The effective Hamiltonian for the metallic layer near its Fermi surface (i.e.  $\xi^{(2)}_{\mvec{k}}\simeq-\xi^{(2)}_{\mvec{k}}$)
 is obtained in a standard way \cite{AssaBook}
\be
h_{2,\bk}^{\rm eff}(E)=h_{2,\mvec{k}} + V^\dag [E-h_{1,\bk}]^{-1} V.\nn\\
\ee
Note that the effective Hamiltonian is energy dependent, and therefore not really a Hamiltonian.
This is because it is  defined through the resolvent operator projected to the lower right block
 \be
 G_{22}(E)=P_2 [E-h]^{-1} P_2 \equiv [E-h_2^{\rm eff}(E)]^{-1}. \label{G22}
 \ee
 The energy dependence will be important below when we consider response to external fields. However, for now, since we are only interested in the low energy spectrum, compared  to the separation between blocks, we may neglect the energy dependence to leading order in degenerate perturbation theory and obtain
\be
h_{2,\bk}^{\rm eff}\!=\!
\begin{pmatrix}
\xi_{2,\mvec{k}}\!-\!\left({\tilde{t}_\perp}/{E_{1\mvec{k}}}\right)^2 \xi_{1,\mvec{k}} &  \left({\tilde{t}_\perp}/{E_{1\mvec{k}}}\right)^2 \Delta_{\mvec{k}} \\
\left({\tilde{t}_\perp}/{E_{1\mvec{k}}}\right)^2\Delta_{\mvec{k}} &\!\!-\xi_{2,-\mvec{k}}\!+\! \left({\tilde{t}_\perp}/{E_{1\mvec{k}}}\right)^2 \xi_{1,-\mvec{k}}
\end{pmatrix}.
\label{h2k}
\ee
Here $E_{1\bk}=\sqrt{\xi_{1\bk}^2+\D_\bk^2}$ is the energy of a quasiparticle of the superconducting layer at a wave vector $\bk$ near the Fermi surface of the normal layer.
We see that a small $t_\perp$ leads to a proximity gap $\Delta_{\mvec{k}}^{\rm prox}= ({\tilde t}_\perp/E_{1\mvec{k}})^2 \Delta_{\mvec{k}}$.
The proximity gap inherits the d-wave symmetry from the pairing gap $\D_\bk$ of the superconducting layer, but is suppressed in magnitude. An important observation is that the degree to which the proximity gap is suppressed is highly sensitive to the Fermi surface matching between the two layers. For highly matched
Fermi surfaces the energy denominator $E_{1\mvec{k}}$ is very small compared to the full bandwidth $4t_1$. In this way it is possible to gain significant enhancement in $T_c$ with relatively small inter-layer tunneling $t_\perp$. It is interesting to note that between a highly overdoped layer with hole concentration $x\sim 0.35$ and an underdoped layer, as seen in angle resolved photoemission (ARPES) experiments, is impressively good\cite{ShenLSCO}. This is not so in the case of gold deposited on the underdoped film, for which no enhancement of $T_c$ was found in Ref.\onlinecite{Yuli}.

In addition to (\ref{h2k}), the complete low energy Hamiltonian also includes the Dirac quasiparticles of the original superconducting layer, which are now slightly renormalized by degenerate perturbation theory (near $E_{1\bk}\simeq-E_{1\bk}$),
 \begin{eqnarray}\label{H1eff}
h_{1,\bk}^{\rm eff}=
\begin{pmatrix}
\xi_{1,\bk}-\tilde{t}_\perp^2/ \xi_{2\bk}& \Delta_\bk\\
\Delta_\bk  &-\xi_{1,-\mvec{k}}+\tilde{t}_\perp^2/ \xi_{2\bk}
\end{pmatrix}.
\nonumber
  \end{eqnarray}

The low energy effective theory captures correctly properties related to quasiparticle excitations at low temperature. To get a full picture of zero temperature properties we should include perturbative corrections to all negative energy states, including those far below the underlying Fermi surfaces of the two layers. Such corrections will be discussed in section \ref{subsec:zeroT}, where we analyze a unique paramagnetic contribution to the stiffness at zero temperature.

\subsection{Response to transverse vector potential}\label{subsec:response}

To compute the superfluid stiffness within the low energy effective theory using formula (\ref{rhos}), we need to derive the renormalized coupling to an external vector potential in
the effective Hamiltonian. This is accomplished by carrying out the renormalization scheme outlined above in the presence of a field, while keeping terms up to second order in $\bf A$ throughout. For simplicity let us take $\bA=A{\hat{\bx}}$.

The coupling of the microscopic Hamiltonian (\ref{Hmatrix}) to the vector potential, up to second order in $A$, is given  by
\be
h_l(\bk,A)= h_l(\bk,0) -\bJ_{l}\cdot\bA + \half K_{l}(\bk)A^2
\label{hA}
\ee
where the index $l=1,2$ refers to the two layers, and $K_{l}(\bk)= \partial_{k_x}^2\xi_l \s_3\equiv \xi_l'' \s_3$ are the kinetic energy operators due to motion along the axis defined by $\bA$ (in our case the $\hat{\bx}$ axis). $\bJ_l=\a_l {\bf v}_{Fl}\s_0$, is the electric current operator on the layer $l$, with $\a_l$ the quasiparticle charge on that layer and ${\bf v}_{Fl}$ the Fermi velocity. The off diagonal block $V$ of the Hamiltonian does not couple to the electromagnetic field.

Now following the same steps as above we can eliminate the coupling between the blocks and
obtain an effective Hamiltonian valid near the Fermi surface of the metallic layer
\be
h_2^{\rm eff}(E,\bA)=h_{2}-\bJ_2 \bA + V^\dag [E-h_{1}+\bJ_1\cdot\bA - \half K_{1}A^2]^{-1} V,
\label{hEA}
\ee
where we have dropped the argument $\bk$ for notational simplicity.
We note that the energy $E$ should be understood as a solution to the equation $\det(E-h_2^{\rm eff}(E))=0$ for the poles of (\ref{G22}). Therefore the energies implicitly depend on the external field $\bA$. To zeroth order in ${\tilde t}_\perp/E_1$ we have $E(\bA)=E(0)-\bJ_2\cdot \bA + \half {\rm sgn}(E(0))\xi_2''A^2$. We must keep the $A$ dependence since we are interested in the response to the external field. However we may still neglect the constant $E(0)$, which is much smaller than $E_1$ in this regime. In this way we obtain the effective Hamiltonian

\begin{widetext}
\bea
h_2^{\rm eff}(\bA)&=&h_2-\bJ_2\cdot\bA+\half K_2 A^2 -V\yd[h_1-(\bJ_1-\bJ_2)\cdot\bA+\half
(\rm sgn(E)\  \xi_2''\s_0-\xi_1''\s_3)A^2]^{-1} V\nn\\
&=& h_2-V\yd h_1^{-1} V\nd-\bJ_2\cdot\bA+\half K_2 A^2 -\left({\tilde{t}_\perp\over E_1}\right)^2 (\bJ_1-\bJ_2)\cdot\bA\nn\\
&&-\left({\tilde{t}_\perp\over E_1}\right)^2 \left(
h_1^{-1}(\bJ_1-\bJ_2)^2-\half\left({\rm sgn}(E)\xi_2''-\xi_1''\s_3\right)\right)A^2.
\label{h2exp}
\eea
\end{widetext}
This is still energy dependent because of the term $\rm sgn(E)$.
However, of the two terms quadratic in $A$ in the last line, the first is larger by a factor  $\sim 4t_1/E_1\sim \pi/\d k$ (see appendix for detailed explanation), where $\d k$ is the mismatch between the Fermi surfaces of the two layers at the nodes. In other words the first term is strongly enhanced by good Fermi surface matching, which is indeed observed by in ARPES experiments done with samples of varying doping levels\cite{ShenLSCO}. Specifically, for a bilayer with underdoped layer at $x=0.07 - 0.15$ and overdoped layer doping $y=0.35$ the estimated ratio is $t_1/E_1\gtrsim 5$. We therefore neglect the energy dependent term and obtain an effective Hamiltonian
\bea
h_2^{\rm eff}(\bA)&=& h_{2,\bk}^{\rm eff}-\bJ_2^{\rm eff}\cdot\bA+\half K_2^{\rm eff} A^2
\label{heff2A}
\eea
with
\bea
\bJ_2^{\rm eff}&=&\bJ_2+\left({{\tilde t}_\perp\over E_1}\right)^2(\bJ_1-\bJ_2)\nn\\
K_2^{\rm eff}&=&K_2-2\left({{\tilde t}_\perp\over E_1}\right)^2\frac{\xi_1\sigma_3+\Delta\sigma_1}{E_1^2} (\bJ_1-\bJ_2)^2.
\label{Jeff2A}
\eea

The effective Hamiltonian $h_1^{\rm eff}(\bA)$ valid near the Dirac nodes of the underdoped layer is derived in the same way,
\bea
h_1^{\rm eff}(\bA)&=& h_{1,\bk}^{\rm eff}-\bJ_1^{\rm eff}\cdot\bA+\half K_1^{\rm eff} A^2
\label{heff1A}
\eea
where
\bea
\bJ_1^{\rm eff}&=&\bJ_1+\left({{\tilde t}_\perp\over \xi_2}\right)^2(\bJ_2-\bJ_1)\nn\\
K_1^{\rm eff}&=&K_1-2{{\tilde t}_\perp^2\over \xi_2^3}\sigma_3 (\bJ_1-\bJ_2)^2.
\label{Jeff1A}
\eea

The phase stiffness we wish to compute can be divided into two parts. First is the zero temperature superfluid stiffness, which to leading order in the inter-layer coupling is given simply by the sum of contributions from the two layers. Second, is the linear reduction of the stiffness with the temperature. Because this reduction is induced by thermal excitation of low energy quasiparticles at the gap nodes, it can be computed using the effective low energy theory. This will be done in the next subsection. We note that there are also zero temperature corrections to the stiffness due to coupling between the layers. These are somewhat more subtle and will be considered in subsection \ref{subsec:zeroT}.
\\

\subsection{Temperature dependent phase stiffness}\label{subsec:Tdepstiffness}
In the low energy effective Hamiltonian we have achieved effective decoupling of the two layers. Therefore, the contributions of the stiffness due to each layer, within this theory, can be added separately and they must each be non negative
\be
\rho_s(T)=\max(\rho_{1},0)+\max(\rho_{2},0).
\label{rhotot}
\ee

At zeroth order in the inter-layer coupling, the zero temperature stiffness is simply the sum of the contributions of the independent layers, $\rho_l(0)=n_{sl}/m_l^\star$.
The leading temperature dependence of $\rho_l$ is a linear reduction in temperature due to a paramagnetic contribution from thermally excited quasiparticles in the nodes of  $h_1^{\rm eff}$ and $h_2^{\rm eff}$. This contribution can be calculated exactly as in Ref. [\onlinecite{WenLee}] using the effective Dirac Hamiltonians $h_l^{\rm eff}$ and the respective quasiparticle currents $J_l^{\rm eff}$:
\be \label{drhopara}
\d \rho_{l,\rm para}= -{8\over T}\sum_{\bk}
\left(J_l^{\rm eff}\right)^2 n_F(\bk)\left(1-n_F(\bk)\right).
\ee
We carry out the integration using the density of states of the respective layers,
\bea
\label{app:DOS}
\nu_1(E)&=&\frac{E}{2\pi v_{F1}v_\Delta}\nn\\
\nu_2(E)&=&\frac{E}{2\pi \tilde{v}_{F2} \tilde{v}_\Delta}
\eea
where $\tilde{v}_\D\equiv v_\D \tilde{t}_\perp^2/E_1^2$ and $ \tilde{v}_{F2}=v_{F2}-v_{F1} \tilde{t}_\perp^2/E_1^2$, the proximity induced gap and Fermi velocities near the Fermi surface of the metallic layer.
Taking $J_l^{\rm eff}$ from Eqs. (\ref{Jeff2A}) and (\ref{Jeff1A}) we  obtain the low temperature contributions to the phase stiffness due to each of the layers
\begin{widetext}
\bea
\rho_{1}&=&\rho_{1}(0)-T{2\ln 2\over \pi}\a_1^2 {v_{F1}\over v_\D}+O\left({\tilde{t}_\perp^2\over \xi_2^2}\right)\nn\\
\rho_{2}&=&\rho_{2}(0)-T{2\ln 2\over \pi}\left( \a_2^2 {v_{F2}\over v_\D}\left({E_1^2\over \tilde{t}_\perp^2}-2\right) + {v_{F1}\over v_\D}(\a_2^2+2\a_1\a_2)\right)
+O\left({\tilde{t}_\perp^2\over E_1^2}\right).
\label{rho1rho2}
\eea
\end{widetext}

Clearly the dominant term in the temperature dependence of $\rho_s$, is due to thermal excitation of quasiparticles in the proximity-induced Dirac cones of the metal (layer 2). This term scales as $v_{F2}/\tilde{v}_{\D}=(v_{F2}/v_{\D})(E_1/{\tilde t}_\perp^2)$.
Terms of order 1 in the dimensionless inter-layer tunneling $ \tilde{t}_\perp/E_1$ are due to excitations in the original Dirac cones and to quasiparticle mixing between the layers.

Our next step is to estimate\cite{footnote} $T_c$ of the bilayer using the formulae (\ref{rhotot})
and (\ref{rho1rho2}). These formulae are expressed mostly in terms of phenomenological parameters, which may in principle be extracted from experiments with bulk samples.

We estimated the needed parameters using the following information:
(i) The zero temperature stiffness of the underdoped layer $\rho_1(0)$ was taken from data interpolation of penetration depth measurements\cite{panagopoulos:2003}. (ii) In the underdoped regime $d\rho_1/dT\approx -1$ almost independent of doping \cite{panagopoulos:1999}. (iii) $v_\Delta$ is determined from the maximal gap extracted from the leading edge shift in ARPES\cite{Ino:2002} by assuming a pure d-wave gap function $\D_\bk=\D(\cos k_x-\cos k_y)$. (iv) The Fermi velocity of the underdoped material is taken from ARPES measurements\cite{ShenLSCO,shen:Nature2003}, which give $\simeq 1.8\ eV-A$ almost independent of the doping within the underdoped regime. (v) The effective quasiparticle charge in the underdoped regime, $\alpha_1(x)= [(\pi/2\ln 2)(v_\D/v_{F1})d\rho_1/dT]^{1/2}$, is then fully determined by (ii)-(iv). (vi) The zero temperature stiffness of the metallic layer is estimated as the average kinetic energy per bond,  $\rho_2(0)=t_2(1-y)$, where we take $y=0.35$. The hopping $t_2\sim 300\ meV$ is taken from the band structure determined by ARPES measurements of LSCO samples\cite{ShenLSCO}. The ratio of the two Fermi velocities is seen to be $v_{F2}/v_{F1}\approx 1.5$.

The one parameter that cannot be extracted from such experiments is the dimensionless inter-layer coupling ${\tilde t}_\perp/E_1$.
We remind the reader that $E_1(\bk)$ is the energy of a quasiparticle of the superconducting layer at the wave-vector $\bk$ near the Fermi surface of the metallic layer. It therefore depends crucially on the distance $\d k$ between the two Fermi surfaces and can be approximated as $E_1\approx \d k v_{F1}$. In principle $\d k$
 may be extracted from ARPES experiments, such as Ref.[\onlinecite{ShenLSCO}]. For underdoped LSCO layer with hole concentration between $x=0.07$ to $x=0.15$, matched with a highly overdoped layer $x=0.35$, we extract $E_1\approx 60\ meV \ll t$. It is therefore possible, in principle, to obtain a sizable proximity effect, even for inter-layer tunneling substantially smaller than $t$ as long as $t_\perp$ is not much smaller than $E_1$.

\begin{figure}[t]
\begin{center}
\includegraphics{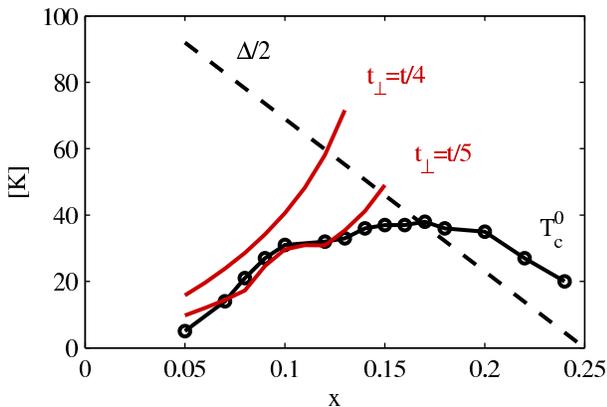}\end{center}
\caption{{\em Phase diagram of bilayer LSCO from the phenomenological theory}. Plots of the critical temperature for a bilayer with different inter-layer tunneling $t_\perp/t$ computed using the phenomenological approach of section \ref{subsec:Tdepstiffness} (solid red curves). This is compared to the measured critical temperature in  bulk LSCO (black circles) taken from Ref. [\onlinecite{panagopoulos:2003}]. The dashed line is a linear interpolation of data for $\Delta(x)/2$ (anti-nodal gap)\cite{Ino:2002}, taken as an estimate for the mean-field critical temperature.
}
\label{fig:phenLSCO}
\end{figure}

Fig. \ref{fig:phenLSCO} shows the phase diagram of LSCO bilayers estimated using the phenomenological theory described in this section.
In the underdoped side $T_c$ is controlled by the temperature dependence of the superfluid stiffness  of the bilayer as given by Eqs. (\ref{rho1rho2}) and (\ref{rhotot}). The value of $T_c$ is determined by the criterion $\rho_s(T_c)=(2/\pi)T_c$ for a Kosterlitz-Thouless transition in the two dimensional interface layer.
The result for $T_c(x)$ of the double layer systems for two values of the inter layer coupling is given by the solid lines in the figure\cite{Footnote:max}.
These lines are cut off by the dashed curve which is the estimate of $T_c$ more appropriate to the overdoped side of the phase diagram. There, the pairing gap becomes smaller than the energy scale set by the superfluid stiffness and therefore the gap sets the scale for $T_c$, which can be estimated by the BCS relation $T_{C}\approx \Delta(x)/2$.  The value of the gap as a function of doping is taken from an interpolation of ARPES data\cite{Ino:2002}. For comparison we also show the transition temperature measured in bulk LSCO\cite{panagopoulos:2003} (black circles).

For inter-layer tunneling $t_\perp\gtrsim t/5$ we see a significant enhancement of  $T_c$ compared to the bulk transition temperature. This is one of our main results. Furthermore the optimal doping level is shifted down compared to the bulk optimal doping, in qualitative agreement with experiment\cite{Yuli}. We point out that for inter-layer tunneling $t_\perp=t/5$ the perturbative parameter is $({\tilde t}_\perp/E_1)^2\sim0.4$, justifying the expansion in (\ref{rho1rho2}). However, we note again that the inter-layer tunneling required to achieve the enhancement of $T_c$ is rather large to directly explain this experiment\cite{Ando_comment,Ando}.

One possible explanation of the experimental result, in-line with our analysis, is that in reality the interface layers share dopants, such that each layer is an inhomogeneous mixture of underdoped  and overdoped puddles.
This is a natural scenario in the samples of Ref. \onlinecite{Yuli}, in which the interface is not atomically sharp and it was shown to consist of facets of the two material components. However inhomogeneous doping of the interface is plausible even in the atomically sharp interfaces of Ref. \onlinecite{BozovicNature}. Indeed these authors mapped the doped hole distribution along the c-axis using resonant X-ray scattering and found that the interface doping is approximately the average of the nominal
doping levels of the two material components\cite{BozovicArxiv}. This was explained by a simple theory of electrostatic screening. The hole distribution within the interface plane was not mapped, but it is highly likely to be inhomogeneous given the random dopant distribution.

 The essential competition of effects that determine $T_c$ in a inhomogeneous layer is expected to be the same as discussed above. The large superfluid density donated by the proximity gapped overdoped regions counters the steep reduction of stiffness with temperature due to the smallness of the proximity gap. Most importantly, now that the two 'phases' are intertwined in the same layer, the effective coupling between them can be much larger. Our analysis of two homogeneous layers with substantial coupling between them could then be viewed as a crude effective description of the inhomogeneous system.

\subsection{Zero temperature paramagnetic response}\label{subsec:zeroT}

So far we were concerned with the variation of the stiffness with temperature. In this section we point out an interesting zero temperature effect of the bilayer coupling on the superfluid stiffness.
Specifically, the superfluid stiffness of the coupled double layer system is smaller than the summed stiffness of the individual layers.

One way to see this effect is by inspection of the effective hamiltonian of the double layer, as given by Eqs.  (\ref{heff2A}) and (\ref{heff1A}). In the effective hamiltonian of each of the layers the quadratic coupling to a vector potential is renormalized down, at second order in the inter-layer tunneling, by a factor proportional to $(\a_1 v_1-\a_2 v_1)^2$. This is not the full contribution to the zero temperature stiffness, which
 consists of the response of the ground state energy to the vector potential. The ground state energy, in turn, involves a sum of all the negative energy solutions of (\ref{Hmatrix}). It is therefore not enough to compute the contribution from the low energy excitations, encoded by the effective Hamiltonians (\ref{heff1A}) and (\ref{heff2A}). As opposed to the calculation of the temperature dependence presented above, here we must also account for the contribution of the negative energy solution of the high energy excitation branch. That is, the energy $E_{1-}(\bk,A)$ at wave-vectors near the Fermi surface of the metallic layer (layer 2) and $E_{2-}(\bk,A)$ near the nodes of the underdoped layer (layer 1).

Thus, the contributions to the superfluid stiffness from the wave-vectors near the fermi surfaces of the two layers are given by:
\be
\rho_\a(0)=\sum_{\bk\in \{\a\}}\left[\xi_1''+\xi_2''+\frac{d^2E_{{\bar\a}-}}{dA^2}+\langle K_\a^{\rm eff} \rangle_{0}\right].
\label{rho02}
\ee
Here the two layers are denoted by $\a=1,2$, while ${\bar \a}=2,1$ denotes the other layer. $K_\a^{\rm eff}$ are given in (\ref{Jeff2A}) and (\ref{Jeff1A}) and $\langle ...\rangle_{0}$ denotes a ground state expectation value. Using the same procedure as outlined in section \ref{subsec:response}, but applied to the large negative energy solutions, we get
\be
\frac{d^2E_{\a-}(\bA)}{d\bA^2}\simeq-\frac{\xi_\a}{E_\a}\xi_\a'' -\frac{\tilde{t}_\perp^2}{E_\a^3}(\bJ_1-\bJ_2)^2.
\label{d2EdA2}
\ee
Finally, using (\ref{rho02}) and (\ref{d2EdA2}) and interpolating to all wave-vectors we obtain the correction to the zero temperature stiffness to order $(\tilde{t}_\perp/E_\a)^2$

\begin{align}\label{T0para}
\delta \rho(0)&\simeq-(\a_1v_{F1}-\a_2v_{F2})^2\sum_{\bk}\frac{2\tilde{t}_\perp^2 \sin^2{k_x}}{(E_{1\bk}+|\xi_{2\bk}|)^3}\\
\nonumber
&\qquad\qquad\qquad\qquad\qquad\times\left(1-\frac{\xi_{1\bk}\xi_{2\bk}}{E_{1\bk}|\xi_{2\bk}|}\right).
\end{align}

This is added of course to the zeroth order stiffness of the two layers $\sim n_{s1}/m^\star_1+n_{s2}/m^\star_2$ (see Eq. \ref{app:rhodia}).

To gain better understanding of the zero temperature paramagnetic correction
and the processes involved, it is worthwhile to derive it from a diagrammatic approach. At second order
in $t_\perp$, the diamagnetic and paramagnetic corrections to the superfluid density are given by the following diagrams
\begin{widetext}
\bea
	\d\rho_{\rm dia}&=&\cbox{\includegraphics{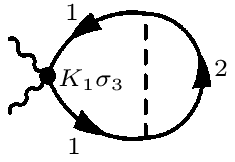}}+\cbox{\includegraphics{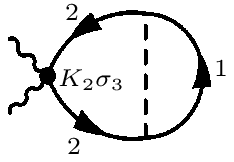}}\nn\\
	\d\rho_{\rm para}&= &+\cbox{\includegraphics{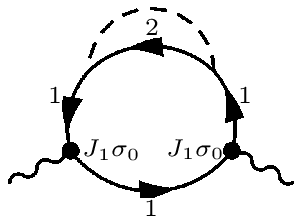}}+\cbox{\includegraphics{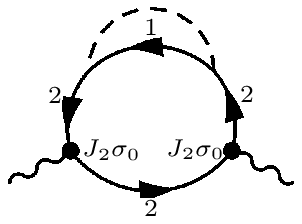}}
	+2~\times~\cbox{\includegraphics{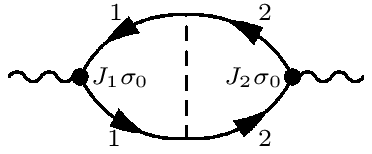}}
\eea
\end{widetext}
Here the labels $1$ and $2$ denote the bare Nambu Green's functions of the isolated layers 1 and 2 respectively and a vertex with a dashed line denotes an inter-layer tunneling process. In adding up the contributions of the  paramagnetic bubble diagrams above, we directly obtain the result (\ref{T0para}). Note that the first two diagrams describe renormalization of the Green's function of each layer due to virtual hopping of electrons out of it into the other layer. The third diagram is a vertex correction describing indirect coupling to the vector potential via hopping to the other layer. The correction to the stiffness of the same order in $t_\perp$ coming from the diamagnetic diagrams is suppressed by a factor of order $\d k /k_F$ . It can therefore be neglected in the case of good Fermi surface matching.

In the appendix we present a more complete derivation of the result (\ref{T0para}) using degenerate perturbation theory. In addition we show there that the paramagnetic correction behaves as $(\a_1 v_1-\a_2 v_2)^2$ at all orders in $t_\perp$.
The same conclusion also emerges from the results of the self consistent mean field calculation (described in section \ref{section:SBMFT}). Fig. \ref{fig:rho0vsJ} shows a quadratic dependence of the zero temperature stiffness on $(J_1-J_2)$ when the effective quasiparticle charge of the underdoped layer is varied.
It is interesting to note the weak dependence of $\rho_s(0)$ on $t_\perp$ at the point $\bJ_1-\bJ_2=0$. This arises from the small diamagnetic term, which we neglected in the analytic calculation.

The paramagnetic correction at zero temperature can be used to measure the current carried by a quasiparticle, and specifically its doping dependence. As mentioned above, this property also comes up in the temperature dependent stiffness, and its dependence on hole doping has posed a long standing puzzle (see for example Ref.  [\onlinecite{LeeReview}]). Since these measures of the quasiparticle charge (or current), are model dependent it is useful to have an independent probe, such as the zero temperature paramagnetic effect in a bilayer.

To define a concrete experiment along these lines it is simpler to consider a bilayer or heterostructure consisting of two types of underdoped layers with a small mismatch in doping $|x-y|\ll x,y$. The experiment involves comparison between the superfluid stiffness measured for the heterostructures to that of the pure materials.
Using the diagrammatic perturbation theory to second order in the inter-layer coupling we obtain the reduction of the zero temperature stiffness in the heterostructure:
\begin{align}\label{deltarhopara:udud}
\delta \rho(0)&\simeq-(\a_1v_{F1}-\a_2v_{F2})^2\sum_{\bk}\frac{2\tilde{t}_\perp^2 \sin^2{k_x}}{(E_{1\bk}+E_{2\bk})^3}\\
\nonumber
&\qquad\qquad\qquad\qquad\times\left(1-\frac{\xi_{1\bk}\xi_{2\bk}-\D_{1\bk}\D_{2\bk}}{E_{1\bk}E_{2\bk}}\right).
\end{align}
Note that here the electron operators of both layers are renormalized and therefore the effective interlayer coupling is $\tilde{t}_\perp=\sqrt{\d(x)\d(y)}t_\perp$. The main contribution to the sum (\ref{deltarhopara:udud}) is from wave-vectors between the underlying fermi surfaces of the two layers. Thus in the limit of good fermi surface matching between the layers ($\d k \ll k_F$) we obtain
\begin{equation}\label{deltarhopara:udud:estimate}
\delta \rho(0)\propto-\frac{xy(\a_1-\a_2)^2}{\d k^2(x,y)}\frac{{t}_\perp^2 k_F}{v_F}.
\end{equation}
where we assumed that in the underdoped regime the fermi velocity is doping independent and thus $v_{F1}=v_{F2}\equiv v_F$. The fermi wave vector depends very weakly on the doping and we denote by $k_F$ the average value of the two layers. In addition we plugged $\d(x)\propto x$ which is valid at low doping levels. If we assume that $\d k \propto x-y$ (as should be expected from the Luttinger theorem), a measurement of the paramagnetic reduction at $T=0$ as function of both $x$ and $y$ can reveal the doping dependence of the effective quasiparticle charge $\a$ in the underdoped regime. Such a measurement will distinguish between the following scenarios: (i) $\a(x)=\a$ independent of doping, in which case there will be no paramagnetic reduction. Note that the small diamagnetic correction that will survive in this case is positive and therefore cannot be mistaken with the paramagnetic correction. (ii) $\a(x) \sim x$ as implied by slave boson mean field theory, which will result in a finite paramagnetic reduction that scales as $x^2$ to leading order in $y-x$. (iii) $\a(x)$ has some other doping dependence, leading to a more involved doping dependence of the paramagnetic response. In general, if $\a(x)$ depends on the doping as $x^\g$ ($\g\neq 0$), then the leading doping dependence of the paramagnetic reduction is $x^{2\g}$.

\section{Summary}\label{sec:summary}
In this paper we showed that significant enhancement of $T_c$ in cuprate heterostructures is possible under realistic conditions and provided a possible explanation for recent measurements on LSCO bilayers by Yuli et. al. [\onlinecite{Yuli}]. Our analysis indicates that the conditions under which such enhancement of $T_c$ can occur depend crucially on the evolution of the superconducting state with underdoping on approaching the Mott insulator. In particular, the effect is sensitive to the way in which the phase stiffness and the current carried by quasiparticles are renormalized as a function of the doping level. Such questions, pertaining to the fundamental nature of superconductivity in the cuprates, are not yet fully resolved, and we proposed that further experiments with cuprate heterostructures can shed new light on these issues.

The essential idea of $T_c$ enhancement in heterostructures\cite{Kivelson} is based on the observation that the pairing scale in the underdoped cuprates is high, and $T_c$ is limited by the low superfluid density in these materials\cite{Uemura,EmeryKivelson}. By inducing a proximity gap in a nearby metallic layer, the large density of charge carriers in that layer is harnessed to the total superfluid response. However because the proximity gap is typically much smaller than the original gap, the reduction of the superfluid density with temperature is also much steeper in the heterostructure. And so, the question whether $T_c$ can in fact be enhanced in this way is more delicate, and sensitive to the nature of superconductivity in the underdoped material.

To address this question we used a microscopic approach based on the slave Boson mean field theory, as well as a semi-phenomenological theory of the doped Mott insulator.
Straightforward slave boson mean field theory showed no enhancement of $T_c$ in the bilayer. Interestingly however, this failure is directly tied to the well known shortcoming of the mean field theory in describing the temperature dependent phase stiffness (see e.g. [\onlinecite{LeeReview}]). One can generalize the low energy theory derived from the microscopic approach to include renormalized parameters for the zero temperature stiffness and the effective charge of a quasiparticle such that it reproduces the observed response in bulk samples\cite{IoffeMillis}. Using such a phenomenological theory for the bilayer we found that $T_c$ enhancement can be achieved for inter-layer tunneling of order $t/5$ or larger.
This value is in excess of the bare inter-layer tunneling in LSCO bilayers, such as those investigated in Ref. [\onlinecite{Yuli}]. We proposed that this discrepancy may be resolved if
each of the layers at the interface is in fact an inhomogeneous mixture of underdoped and overdoped material (e.g. as a result of dopant migration). In this case our bilayer model with moderate coupling $t_\perp \sim t/5$ can be viewed as a crude effective description of the inhomogeneous interface.

We note that the analysis performed in this paper uses a completely homogeneous model. It does not include for example stripe or density waves structures. The existence of such structures therefore does not appear to be crucial for obtaining an enhanced $T_c$. From this point of view the fact that the maximal enhancement seen in Ref. [\onlinecite{Yuli}] was close to 1/8 may be coincidental.

Finally, we pointed out a unique paramagnetic contribution to the zero temperature phase stiffness of a bilayer system. The paramagnetic reduction of the zero temperature stiffness is proportional to $t_\perp^2$ and to $(\bJ_1-\bJ_2)^2$, that is, the square of the difference of electric current carried by a quasiparticle on each layer. We proposed that experiments with bilayers or heterostructures can serve as a new kind of probe of the effective quasiparticle and its doping dependence in  cuprates.

\section{Acknowledgements}
We thank Erez Berg, Sebastian Huber, Amit Kanigel, Steve Kivelson, Oded Millo, Dror Orgad, and Ofer Yuli for stimulating discussions. This work was supported by grants from the Israeli Science Foundation and the Minerva foundation.
\appendix

\section{Perturbation theory in the inter-layer coupling}

Here we use straightforward perturbation theory of (\ref{Hmatrix}) in the inter-layer tunneling to compute the zero temperature stiffness of the bilayer. This is an alternative to the effective Hamiltonian approach used in section \ref{subsec:zeroT} to obtain the zero temperature stiffness and provides a check of the results.
The expansion is separated to different regions in the Brillouin zone where different sets of levels may be nearly degenerate. For example we describe the expansion for wave-vectors near the Fermi surface of the metallic layer (layer 2).  The (non-normalized) eigenvectors of (\ref{Hmatrix}) corrected to first order in $t_\perp/E_1$ and to lowest order in $E_2/E_1$ are given by:
\begin{equation}
\begin{array}{lclccccl}
| 1\rangle &= &\bigl(& u  &   v& \frac{\tilde{t}_\perp}{E_1}u &  -\frac{\tilde{t}_\perp}{E_1}v&\bigr)^{\rm T}\\[6pt]
| 2\rangle &=&\bigl(&  -v & u& \frac{\tilde{t}_\perp}{E_1}v  & \frac{\tilde{t}_\perp}{E_1}u&\bigr)^{\rm T}\\[6pt]
| 3\rangle &=&\bigl(&  \frac{\tilde{t}_\perp}{E_1}{f}_1 &  -\frac{\tilde{t}_\perp}{E_1}{f}_2 &  \bar{u} &\bar{v}&\bigr)^{\rm T}\\[6pt]
| 4\rangle &=&\bigl(&\frac{\tilde{t}_\perp}{E_1}{f}_2 	&\frac{\tilde{t}_\perp}{E_1}{f}_1 &- \bar{v} & \bar{u}&\bigr)^{\rm T}.
\end{array}
\end{equation}
Note that the subscripts $\bk$ of $E_1$, $u$, and $v$ are suppressed for notational simplicity. Here
$u=[(1+\xi_1/E_1)/2]^{1/2}$, $v=[(1-\xi_1/E_1)/2]^{1/2}$, $\bar{u}=[(1+\tilde{\xi}_2/E_2)/2]^{1/2}$, $\bar{v}=[(1-\tilde{\xi}_2/E_2)/2]^{1/2}$, where $\tilde{\xi}_2=\xi_2-\xi_1{\tilde{t}_\perp^2}/{E_1^2}$ and $E_2=[\tilde{\xi}_2^2+\Delta^2{\tilde{t}_\perp^4}/{E_1^4}]^{1/2}$. In addition we denote $f_1\equiv 2uv\bar{v}-(u^2-v^2)\bar{u}$ and $f_2\equiv 2uv\bar{u}+(u^2-v^2)\bar{v}$.

These states can now be used to compute the response to an external vector potential coupled to the Hamiltonian (\ref{Hmf}).
\begin{equation}\label{Hmf:wA}
H_{MF}(A)= \sum_{\mvec{k}}[\frac{1}{2}(\xi_{1\mvec{k}}''+\xi_{2\mvec{k}}'')A^2  + \Psi_\mvec{k}^\dagger h_\mvec{k}(A) \Psi_\mvec{k}]\nn
\end{equation}
where
\be
h_{\bk}(A)=h_{\bk}(0)-
\hat{J}A
+\frac{1}{2}
\hat{K}A^2
\ee
with $h_{\bk}(0)$ given by (\ref{Hmatrix}) and
\be\label{JmatKmat}
\hat{J}=
\begin{pmatrix}
J_1\s_0 & 0\\
0 & J_2\s_0
\end{pmatrix}
\qquad\qquad
\hat{K}=
\begin{pmatrix}
 \xi_{1}''\s_3    &  0  \\
   0   &  \xi_{2}''\s_3
\end{pmatrix}.
\ee

In particular the diamagnetic response at $T=0$ includes the first order correction to the ground state energy in the quadratic coupling term,
\begin{widetext}
\bea\label{app:rhodia}
\rho_{\rm dia}
&=&
\sum_\mvec{k}
\left[
\xi_{1\mvec{k}}''+\xi_{2\mvec{k}}'' + \bra{2}\hat{K}\ket{2}+\bra{4}\hat{K}\ket{4}
\right]
\\
\nonumber
&\simeq&
\sum_\mvec{k}
\left[
\xi_{1\mvec{k}}''
\left(
1-\frac{\xi_{1\bk}}{E_{1\bk}}
\right)
+  \xi_{2\mvec{k}}''
\left(
1-\frac{\xi_{2\bk}}{|\xi_{2\bk}|}
\right)
\right]
+
\sum_\mvec{k}
\frac{\tilde{t}_\perp^2}{E_{1\bk}^2}
\Biggl[
 \xi_{1\mvec{k}}''
 \biggl(
 \frac{\xi_{1\mvec{k}}}{E_{1\bk}}-\frac{\xi_{2\mvec{k}}}{|\xi_{2\mvec{k}}|}\frac{(\xi_{1\bk}^2-\D_{1\bk}^2)}{E_{1\bk}^2}
 \biggr)
-
 \xi_{2\mvec{k}}''
 \biggl(
 \frac{\xi_{1\mvec{k}}}{E_{1\bk}}-\frac{\xi_{2\mvec{k}}}{|\xi_{2\mvec{k}}|}
 \biggr)
\Biggr]
\eea

The zero temperature paramagnetic contribution is given by the second order perturbation theory in the linear coupling term:

\be\label{app:rhopara}
\rho_{\bf{para}}\simeq-2\sum_\mvec{k}\left[ \frac{|\bra{2}\hat{J}\ket{3}|^2}{E_{1\bk}+|\xi_{2\bk}|} +  \frac{|\bra{4}\hat{J}\ket{1}|^2}{E_{1\bk}+|\xi_{2\bk}|} \right]
				 =-2(\a_1v_{F1}-\a_2v_{F2})^2\sum_\mvec{k}\frac{\tilde{t}_\perp^2\sin^2{k_x}}{(E_{1\bk}+|\xi_{2\bk}|)^3}\left( 1-\frac{\xi_{1\bk}\xi_{2\bk}}{E_{1\bk}|\xi_{2\bk}|} \right)
\ee
\end{widetext}
The perturbative corrections to the zero temperature stiffness to order $\tau^2\equiv{\tilde{t}_\perp^2}/{E_{1\bk}^2}$ consist of two contributions. The diamagnetic contribution, proportional to $2t\tau^2$ (with $t=t_1=t_2$) and the paramagnetic contribution, proportional to $2(2t)^2\tau^2/E_1$ and thus larger by a factor of $\approx4t/E_1\sim k_F/\d k$ ($\d k$ is the fermi surface mismatch of the two layers). This consideration allowed us to keep only the perturbative terms $\sim(J_1-J_2)^2$ in the derivation of the effective Hamiltonians (\ref{heff2A}) and (\ref{heff1A}).

The zero temperature paramagnetic response scales as $(J_1-J_2)^2$ to all orders in $t_\perp$. To see this we denote the exact set of four eigenvectors of the Hamiltonian (\ref{Hmatrix}) by $\{\ket{n}\}$ and their corresponding energies by $\mathcal{E}_n$.
We note that the matrix $\hat{J}$ can be rewritten as $(J_1-J_2)\tilde{\s}+J_2 I$ where $I$ is the 4x4 unit matrix and
\be
\tilde{\s}=
\begin{pmatrix}
 \s_0 & 0  \\
0 &  0
 \end{pmatrix}.
\ee
The paramagnetic response is given by the second order correction of the ground state energy in the presence of an external field (as in equation (\ref{app:rhopara})), and thus involves only off-diagonal matrix elements of $\hat{J}$ in the basis $\{\ket{n}\}$. As a result, matrix elements of
$J_2 I$ vanish due to orthogonality of the eigenvectors and we are left with
\be
 \d\rho_{\rm para}=2\sum_\mvec{k}(J_1-J_2)^2\sum_{\substack{n\in {\rm neg}\\m\neq n}}\frac{|\bra{n}\tilde{J}\ket{m}|^2}{\mathcal{E}_n-\mathcal{E}_m}
\ee
where $n\in {\rm neg}$ denotes the negative energy eigenstates.


\begin{thebibliography}{99}
\bibitem{Uemura} Y. J. Uemura \emph{et. al.},
Phys. Rev. Lett. \textbf{62}, 2317 (1989).
\bibitem{EmeryKivelson} V. J. Emery and S. A. Kivelson, Nature \textbf{374},
434 (2002).
\bibitem{Corson} J. Corson, R. Mallozzi, J. Orenstein, N. Eckstein, and I.
Bozovic, Nature (London) \textbf{398}, 221 (1999).
\bibitem{Ong} Y. Wang, L. Li, and N. P. Ong, Phys. Rev. B \textbf{73},
024510 (2006).
\bibitem{Kivelson} S. A. Kivelson, Physica B \textbf{11}, 61 (2002)
\bibitem{Erez}  E. Berg, D. Orgad, S. A. Kivelson, Phys. Rev. B {\bf 78}, 094509 (2008)

\bibitem{Maier}  S. Okamoto, T. A. Maier, Phys. Rev. Lett, {\bf 101}, 156401 (2008).
\bibitem{BozovicPhysica} G. Logvenov, V. V. Butkoa, C. DevilleCavellinb,
J. Seoc, A. Gozar and I. Bozovic, Physica B
{\bf 403}, 1149 (2008).
\bibitem{BozovicNature} A. Gozar, G. Logvenov, L. Fitting Kourkoutis, A. T. Bollinger, L. A. Giannuzzi, D. A. Muller, I. Bozovic, Nature {\bf 455}, 782 - 785 (2008)
\bibitem{Yuli} O. Yuli, I. Asulin, O. Millo and D.
Orgad, L. Iomin, G. Koren, Phys. Rev. Lett. {\bf 101}, 057005 (2008)

\bibitem{WenLee} P. A. Lee and X.-G. Wen, Phys. Rev. Lett. {\bf 78}, 4111 (1997).
\bibitem{Millis} A. J. Millis, S. Girvin, L. Ioffe, and A. Larkin,  J. Phys.
and Chem. of Solids {\bf 59}, 1742 (1998).
\bibitem{IoffeMillis} L. B. Ioffe and A. J. Millis, J. Phys. Chem. Solids {\bf 63},
2259 (2002).
\bibitem{SBMFT1} A. E. Ruckenstein, P. J. Hirschfeld, and J. Appel,
Phys. Rev. B {\bf 36}, 857 (1987)
\bibitem{SBMFT2} G. Kotliar and J. Liu, Phys. Rev. B {\bf 38}, 5142 (1988)
\bibitem{SBMFT3}Y.Suzumura,  Y. Hasegawa, and H. Fukuyama,  J. Phys.
Soc. Jpn. {\bf 57}, 2768 (1988).
\bibitem{LeeReview} P. A. Lee, X. -G. Wen, and N. Nagaosa, Rev. Mod. Phys. {\bf 78}, 17 (2006)
\bibitem{Lemberger} B. R. Boyce, J. Skinta, and T. R. Lemberger, Physica C
341-348, 561 (2000); J. Stajic, A. Iyengar, K. Levin, B. R. Boyce, and T. R. Lemberger,
Phys. Rev. B 68, 024520 (2003).
\bibitem{panagopoulos:1999} C. Panagopoulos, B. D. Rainford, J. R. Cooper, W. Lo, J.
L. Tallon, J. W. Loram, J. Betouras, Y. S. Wang, and C.
W. Chu, Phys. Rev. B {\bf 60}, 14617 (1999).
\bibitem{AssaBook} A. Auerbach, Interacting Electrons and Quantum Magnetism (Springer, Berlin, 1994), chapter 3.
\bibitem{ShenLSCO}T. Yoshida, X. J. Zhou, D. H. Lu, Seiki Komiya, Yoichi Ando, H. Eisaki,
T. Kakeshita, S. Uchida, Z. Hussain, Z.-X. Shen, and A. Fujimori, J. Phys.: Condens. Matter {\bf 19}, 125209 (2007).
\bibitem{footnote} Here we neglect nonlinear corrections to the temperature dependence of the superfluid stiffness which tend to slightly suppress $T_c$.
\bibitem{Footnote:max} Note the qualitative difference between the lines corresponding to $t_\perp=t/5$ and  $t_\perp=t/4$. For a range of doping levels ($x<0.12$) $t_\perp=t/5$ happens to be just under a critical coupling strength below which the metallic layer does not contribute to $T_c$. That is the superfluid density of the metallic layer falls to zero at a temperature lower than the homogeneous underdoped $T_c$. In this regime the curve $T_c(x)$ simply follows the bulk $T_c$ of the underdoped film.
\bibitem{panagopoulos:2003} C. Panagopoulos, T. Xiang, W. Anukool, J. R. Cooper,
Y. S. Wang, and C. W. Chu, Phys. Rev. B {\bf 67}, 220502(R)
(2003).
 \bibitem{Ino:2002} A. Ino, C. Kim, M. Nakamura, T. Yoshida, T. Mizokawa,
A. Fujimori, Z.-X. Shen, T. Kakeshita,H. Eisaki, and S.
Uchida, Phys. Rev. B {\bf 65}, 094504 (2002).
\bibitem{shen:Nature2003} X. J. Zhou et. al., Nature (London) \textbf{423}, 398 (2003).
\bibitem{BozovicArxiv}  S. Smadici, J. C. T. Lee, S. Wang, P. Abbamonte, A. Gozar, G. Logvenov, C. D. Cavellin, I. Bozovic,  arXiv:0805.3189 (preprint).

\bibitem{Ando_comment} The fact that the inter-layer hopping is much smaller than $t/5$ can be inferred, for example, from resistivity measurements\cite{Ando}, which find the anisotropy $\rho_c/\rho_{ab}\sim 1000$.
    \bibitem{Ando} S. Komiya, Y. Ando, X. F. Sun, A. N. Lavrov, Phys. Rev. B 65, 214535 (2002).
\end{thebibliography}
\end{document}